\documentclass[twocolumn,showpacs,unsortedaddress,superscriptaddress,showkeys,preprintnumbers,letterpaper]{revtex4-1}
\usepackage{acronym}
\usepackage{amsmath}
\usepackage{amssymb}
\usepackage{graphics}
\usepackage{subfigure}
\usepackage{comment}
\usepackage[usenames]{color}
\usepackage[normalem]{ulem}

\newcommand{\CITA}{Canadian Institute for Theoretical Astrophysics, 60 St.
George Street, University of Toronto, Toronto, ON M5S 3H8, Canada}
\newcommand{\Perimeter}{Perimeter Institute for Theoretical Physics, Waterloo,
Ontario N2L 2Y5, Canada}
\newcommand{\AEI}{Albert-Einstein-Institut, Max-Planck-Institut f\"ur
Gravitationsphysik, D-30167 Hannover, Germany}
\newcommand{\Leibniz}{Leibniz Universit\"at Hannover, D-30167 Hannover, Germany}

\newcommand*{\ee}{\mathrm{e}}

\newcommand*{\diff}{\,\mathrm{d}}

\newcommand{\snr}[1]{\ensuremath{\rho_{#1}}}

\newcommand{\chisq}[1]{\ensuremath{\chi^2_{#1}}}
\newcommand{\likely}{\mathcal{L}}
\newcommand{\lstar}{\likely^{*}}
\newcommand{\intrinsic}{\ensuremath{\bar{\theta}}}

\newcommand{\trials}{\ensuremath{M}}
\begin{document}

%%%%%%%%%%%%%%%%%%%%%%%%%%%%%%%%%%%%%%%%%%%%%%%%%%%%%%%%%%%%%%%%%%%%%%%%%%%%%%%
% TITLE, AUTHOR
%%%%%%%%%%%%%%%%%%%%%%%%%%%%%%%%%%%%%%%%%%%%%%%%%%%%%%%%%%%%%%%%%%%%%%%%%%%%%%%

\title{A method to estimate the significance of coincident gravitational-wave
observations from compact binary coalescence}

\author{Kipp~Cannon}
\email{kipp.cannon@ligo.org}
\affiliation{\CITA}

\author{Chad~Hanna}  
\email{chad.hanna@ligo.org}
\affiliation{\Perimeter}

\author{Drew~Keppel}  
\email{drew.keppel@ligo.org}
\affiliation{\AEI}
\affiliation{\Leibniz}

%\author{Others}  
%\email{none}
%\affiliation{none}

\begin{abstract}
Coalescing compact binary systems consisting of neutron stars and/or black
holes should be detectable with upcoming advanced gravitational-wave detectors
such as LIGO, Virgo, GEO and {KAGRA}.  Gravitational-wave experiments to date
have been riddled with non-Gaussian, non-stationary noise that makes it
challenging to ascertain the significance of an event.  A popular method to
estimate significance is to time shift the events collected between detectors
in order to establish a false coincidence rate.  Here we propose a method for
estimating the false alarm probability of events using variables commonly
available to search candidates that does not rely on explicitly time shifting
the events while still capturing the non-Gaussianity of the data.  We present a
method for establishing a statistical detection of events in the case where
several silver-plated (3--5$\sigma$) events exist but not necessarily any
gold-plated ($>5\sigma$) events.  We use LIGO data and a simulated, realistic,
blind signal population to test our method.
\end{abstract}

\maketitle
\acrodef{ASD}{amplitude spectral density}
\acrodef{BNS}{binary neutron star}
\acrodef{CBC}{compact binary coalescence}
\acrodef{FAR}{false alarm rate}
\acrodef{FAP}{false alarm probability}
\acrodef{GW}{gravitational-wave}
\acrodef{GRB}{gamma-ray burst}
\acrodef{IMR}{inspiral-merger-ringdown}
\acrodef{PN}{post-Newtonian}
\acrodef{ROC}{Receiver Operator Characteristic}
\acrodef{SNR}{signal-to-noise ratio}
\acrodef{SPA}{stationary phase approximation}
\acrodef{SVD}{singular value decomposition}

%%%%%%%%%%%%%%%%%%%%%%
\section{Introduction}

Detecting the \acp{GW} from coalescing neutron stars and or black holes should be
possible with advanced \ac{GW} detectors such as LIGO, Virgo, GEO and
KAGRA~\cite{rates2010}.  If the performance of past detectors is any indicator
of the performance of future \ac{GW} detectors, they are likely to be affected
by non-Gaussian noise~\cite{slutsky2010}.  Coincident observations are crucial
in validating the detection of \acp{GW} but it is necessary to establish the
probability that the coincident event could arise from noise alone.  

If the detectors' data were Gaussian and stationary, it would be
straightforward to compute the \ac{FAP} of a coincident event based solely on
its \ac{SNR} and the number of independent trials.  With non-stationary,
non-Gaussian data the \ac{SNR} is not sufficient to describe the significance
of an event and, furthermore, the distribution of detector noise is not known a
priori.  

Estimating false-coincident backgrounds from time delay coincidence associated
with searches for \acp{GW} was first proposed for targeted compact binary coalescence
\ac{GW} searches in~\cite{CBC:S1}.  This method has been the commonest used in
subsequent searches~\cite{CBC:S2, LIGO:S2MACHO, LIGO:S2BBH, LIGO:TAMA,
LIGO:S3Spin, CBC:S3S4, CBC:S51yr, CBC:S5:12_18, CBC:S5LV, CBCHM:S5, CBC:S6}.
We present a method to estimate the false alarm probability of a \ac{GW} event
from coalescing compact objects without time shifts by measuring the false alarm
probability distributions for non-coincident events using a set of common
variables available to the searches.  This greatly simplifies analysis and
lends itself nicely to an online analysis environment.

This paper is organized as follows.  In Sec.~\ref{sec:Method} we describe a
formalism for ranking \ac{GW} events and establishing the probability
distribution for a given event's rank in noise.  In Sec.~\ref{sec:pop} we
present how to estimate the significance of a population of CBC events, which
might include silver-plated (i.e.\ less than $5\sigma$) events.  In
Sec.~\ref{sec:examples}, we test our method with a mock, advanced detector
search that uses four days of LIGO fifth science run (S5) data that has been
recolored to have an Advanced LIGO spectrum containing a plausible, simulated,
blind population of double neutron star binary mergers.  We demonstrate that we
can detect  \acp{GW} from neutron star binaries with very low false alarm
probability.  
%We also demonstrate that we can make a gold-plated detection of a
%population of compact binary coalescences without any individual event in that
%population being gold-plated.
%We anticipate this as a likely scenario for the first direct detection of \acp{GW}.

\section{Method}
\label{sec:Method}

\ac{GW} searches for compact binary coalescence begin by matched filtering data
in the detectors~\cite{findchirppaper}.  If peaks in \ac{SNR} times series for
more than one detector are consistent with the light travel time between
detectors and timing errors, these peaks are considered to be a coincident
event.
 
\ac{GW} data to date have not been stationary and Gaussian~\cite{slutsky2010}
thus making it difficult to model the noise in \ac{GW} searches.
Non-stationary noise degrades the effectiveness of standard matched filter
searches.  For that reason additional signal consistency tests are often
employed, such as explicit $\chi^2$ tests~\cite{Allen:2004, HannaThesis}.
Non-stationarity occurs on several timescales.  Here we are more concerned with
short duration non-stationary bursts of noise called glitches for which
$\chi^2$ tests are very useful discriminators.  

In this section we will present a method using common variables available to a
compact binary search to estimate the \ac{FAP} without relying on time shifting
the detector data.  Although many variables and measurements may be used, in
this paper we consider two parameters: the matched filter SNR \snr{i} and the
$\chi^2$ statistic \chisq{i}, which depend on the detector $i$, as well as
parameters intrinsic to the source that the template describes such as mass and
spin, \intrinsic{}.  In this section, we introduce the framework for evaluating
the \ac{FAP} of \ac{GW} candidates.

\subsection{Ranking events}

Here is our concise definition of a coincident gravitational wave search for
compact binary sources.
i) The search consists of $D$ detectors.
ii) We seek to find the significance of an event found in the D detectors
localized in time.
iii) The intrinsic parameters of the event will be unknown a priori.  Our
detection pipeline will measure the significance as a function of the
parameters of the template waveform \intrinsic{}

For each detector $i$ of a $D$ detector network we use \snr{i} and \chisq{i}
to rank candidates with parameters \intrinsic{} from least likely to be a
gravitational wave to most likely.  We use a standard likelihood
ratio~\cite{Gregory:2005} defined as
\begin{equation}
%\begin{multline}
%
\label{eq:rank}
\likely(\snr{1},\chisq{1},\dots \snr{D},\chisq{D},\intrinsic{})
= \frac{P(\snr{1}, \chisq{1}, \dots \snr{D},\chisq{D},\intrinsic{}|s)}{P(\snr{1}, \chisq{1}, \dots \snr{D},\chisq{D},\intrinsic{}|n)},
\end{equation}
%\end{multline}
%
%
where $P(\dots|s)$ is the probability of observing $(\dots)$ given a signal,
and $P(\dots|n)$ is the probability of observing $(\dots)$ given noise.  It is
assumed that the signal distribution has been marginalized over all relevant
parameters and the \intrinsic{} refers only to the template waveform parameters
that are measured by the pipeline.  We make the simplifying
assumption~\cite{Cannon2008} that the likelihood can be factored into products
of likelihoods from individual detectors,
\begin{equation}
\label{eq:LiIndep}
\likely(\snr{1},\chisq{1},\dots \snr{D},\chisq{D},\intrinsic{})
	\approx  \prod_i^D \likely_{i}(\snr{i},\chisq{i},\intrinsic{}).
\end{equation}
The simplification that the likelihood function can be built from these
products implies statistical independence between detectors for both signals
and noise.  This results in a suboptimal ranking statistic.  However, we can
compute the \ac{FAP} associated with this statistic, and in fact, it becomes much
easier to do so.

\begin{comment}
\subsection{Computing the probability that the data contains noise}

Given that we have a formalism for ranking coincident gravitational-wave
candidates via \eqref{eq:rank}, we now need to establish with what probability
we would expect that the data contains only noise given the value of likelihood
and template parameters \intrinsic{}.  Thus we seek $P(n|\likely)$.
Unfortunately we do not have direct knowledge of this probability.  We note
that from Bayes' theorem we have
%
%
\begin{equation}
%
P(n|\likely) = \frac{P(\likely|n) P(n)}{P(\likely)}.
%
\end{equation}
%
% 
Furthermore we note that the distribution of $\likely$ comes from two exclusive
possibilities: i) either from signal $s$ or ii) from noise $n$.
%
%
\begin{equation}
%
P(\likely) = P(\likely|n) P(n) + P(\likely|s) P(s)
%
\end{equation}
%
%
This leads to
%
%
\begin{multline}
%
P(n|\likely) \\
	= \frac{P(\likely|n)P(n)}{P(\likely|n)P(n) + P(\likely|s)P(s)} \\
	= \frac{P(\likely|n)}{P(\likely|n) + P(\likely|s)P(s)/P(n)} \\
	\leq P(\likely|n) \iff P(s) / P(n) \geq \frac{1 - P(\likely|n)}{P(\likely|s)}
%
\end{multline}
%
%
Thus we can bound the probability that the data is noise by using the
probability of getting a certain value of $\likely$ only under the condition
above.  It should be possible to compute the probability $P(\likely|s)$ and
get some handle on the requirements for the prior probabilities.  However, we
leave that as future work.  The rest of this paper will describe the
computation of $P(\likely|n)$, the standard definition of \ac{FAP}.
\end{comment}

\subsection{Computing the \ac{FAP}}

The \ac{FAP} is the probability of measuring a given $\likely$ if the data
contains only noise.  N.B., this is not the same as assessing the probability
that the data contains only noise, which requires knowing the prior
probabilities of both signal and noise.  In constructing the \ac{FAP},
$P(\likely|n)$, we start with
\begin{equation} \label{eq:outer}
P(\likely, \intrinsic{}|n) = \int_{\Sigma} P(\likely_{1}, \dots \likely_{D},
\intrinsic{}|n) \diff^{D-1} \Sigma,
\end{equation}
where $\Sigma$ is the surface of constant $\likely = \prod_{i}^{D}
\likely_{i}$.  From \eqref{eq:LiIndep}, we have, assuming that the likelihood
values in noise are independent between the detectors, 
\begin{equation} \label{eq:pn}
P(\likely_{i}, \dots \likely_{D}, \intrinsic{}|n) = \prod_{i}^D P(\likely_{i},
\intrinsic{}|n),
\end{equation}
where $P(\likely_{i}, \intrinsic{}|n)$ is obtained by marginalizing over
\snr{i}, and \chisq{i} in the single-detector terms,
\begin{equation} \label{eq:pnmarg}
P(\likely_{i}, \intrinsic{}|n) = \int_{\sigma}
P(\snr{i},\chisq{i},\intrinsic{}|n) \diff \sigma,
\end{equation}
where $\sigma$ is the contour of constant $\likely_{i}$ in the \{\snr{i},
\chisq{i}\} surface at constant \intrinsic{}.  Implicit in \eqref{eq:pn} and
\eqref{eq:pnmarg} is the assumption that the coincidence criteria do not depend
on \snr{i}, \chisq{i} or \intrinsic{}.  Finally, $P(\likely|n)$ is obtained by
marginalizing over $\intrinsic{}$, \begin{equation} P(\likely|n) = \int
P(\likely, \intrinsic{}|n) \diff \intrinsic{}.  \end{equation}

The probability of observing an event with a likelihood value at least as
large as some threshold $\lstar$ is
\begin{equation} \label{eq:cumdist}
P(\lstar|n) 
	:= P(\likely \geq \lstar|n)
	= \int_{\lstar}^{\infty} P(\likely|n) \diff \likely.
\end{equation}
A \ac{GW} search will typically produce multiple coincident events during a
given experiment.  That means that there will be multiple opportunities to
produce an event with a certain likelihood value.  We are ultimately interested
in the probability of getting one or more events with $\likely$ $\geq$ $\lstar$
after all the events are considered.  The probability of getting at least one
such event after forming \trials{} independent 
% Foot note
coincidences\footnote{In practice it can be difficult to know if the
coincidences formed are independent, however as long as they are related to the
true number of independent trials by an overall scaling, one can adjust the
number so that it agrees with the observed rate of coincidences for low
significance events. This works because \acp{GW} are very rare and true signals
will vastly underwhelm the false positives that a pipeline produces at high
\ac{FAP}. Thus the bias in calibrating \trials{} to high \ac{FAP} events is
very small.}
can be adjusted by the complement of the binomial distribution
\begin{multline}\label{eq:PlstarM}
P(\lstar|n_1,\ldots,n_{\trials{}}) :=
	1 - \binom{\trials{}}{0}P(\lstar|n)^0 (1-P(\lstar|n))^{\trials{}} \\
	= 1 - (1-P(\lstar|n))^{\trials{}}.
\end{multline}
This is the \ac{FAP} at $\lstar$ in an experiment that yielded $M$ coincident
events. In what follows, we will drop the explicit $n_1,\ldots,n_{\trials{}}$
notation and simply use $n$ where it is assumed that we have corrected for the
number of trials.

\section{\ac{GW} Events as a Poisson Distribution}
\label{sec:pop}

Historically, \ac{GW} experiments have used rates to rank
events~\cite{CBC:S51yr, CBC:S5:12_18, CBC:S5LV, CBCHM:S5, CBC:S6}.  Assuming
that the likelihood function is independent of time over the duration of the
experiment (or can be approximated as such) we can cluster the most significant
events of the search over a duration longer than the correlation induced by the
filter and we might expect the events arising from noise to obey Poisson
statistics.  In what follows we assume that in fact this is the case and
connect our estimation of \ac{FAP} with the \ac{FAR} often quoted in
gravitational wave searches for compact binary coalescence.

For a Poisson process with mean $\lambda$, the probability of observing $N$ or
more events is given by the survival function
\begin{equation} \label{eq:poissonN}
P(N|\lambda) = 1 - e^{-\lambda} \sum_{i=0}^{N-1}  \frac{\lambda^i}{i!}.
\end{equation}
Using $P(\lstar)$ from \eqref{eq:PlstarM}, setting $N=1$, and solving for
$\lambda$ tells us the mean number of noise events with $\likely\geq\lstar$,
\begin{equation} \label{eq:poissonrate}
\lambda(\lstar) = -\ln \left[ 1 - P(\lstar) \right].
\end{equation}
The quantity, inverse false alarm rate, is given by $\mathrm{IFAR} =
T/\lambda$, where $T$ is the observation time of the experiment.

If $\lstar_{N}$ is the likelihood of the $N^{\mathrm{th}}$ most significant
event, the number of background events expected with $\likely \geq \lstar_{N}$
is
\begin{equation} \label{eq:poissonlstar}
\lambda(\lstar_{N}) = -\ln \left[ 1 - P(\lstar_{N}) \right].
\end{equation}
Since we observed $N$ events with $\likely \geq \lstar_{N}$, the probability of
having produced at least this many events is found by substituting
\eqref{eq:poissonlstar} into \eqref{eq:poissonN},
\begin{equation} \label{eq:poissonNlstar}
P(N|\lambda(\lstar_{N})) = 1 - e^{-\lambda(\lstar_{N})} \sum_{i=0}^{N-1}
\frac{\lambda(\lstar_{N})^i}{i!}.
\end{equation}
A population of events can collectively be more significant than the single
most significant event alone. Indeed, population analyses have previously been
employed in looking for \ac{GW} signals associated with \acp{GRB}. For example,
a Student-T test was proposed in~\cite{FinnGWGRB1999} to test for deviations in
the cross-correlation of detectors' output preceding a set of times associated
with \acp{GRB} (i.e., \emph{on-source} times) when compared to other
\emph{off-source} times not associated with \acp{GRB}, a binomial test was
employed in~\cite{LIGO:Burst:S2-S4:GRB, LIGO:Burst:S5:GRB} using the X\% most
significant events to test for excess numbers of events at their associated
\acp{FAP}, a Kolmogorov test was used in~\cite{Bars:GRB:2005} to look for
deviations from isotropy in \ac{GRB} direction based on the directional
sensitivity of the bar detectors, and a Mann-Whitney U (or
Mann-Whitney-Wilcoxon) test was performed in~\cite{LIGO:CBC:S5:GRB} to test if
the all the \acp{FAP} associated with the on-source events of the \acp{GRB}
were on average smaller than the expected distribution given by the off-source
events, as would be the case if the average significance were elevated due to
the presence of \acp{GW} in the on-source events.

As noted in~\cite{LIGO:Burst:S2-S4:GRB, LIGO:Burst:S5:GRB}, seeking
significance by considering different choices of population diminishes the
significance of each on account of the trials that have been conducted. We
control this by restricting ourselves to considering only populations
consisting of contiguous sets of events that include the most significant, and
are limited to a maximum size $N_{\max}$ where $N_{\max}$ is the rank of the
most significant event at whose ranking statistic (IFAR) value the expected
number of background events was greater than 1.
% See App.~\ref{app:background1} for why we consider this a
%reasonable stopping condition.
There are $N_{\max}$ choices of population possible, so we incur that cost from
the number of trials, modifying the \ac{FAP} in \eqref{eq:poissonNlstar} to
\begin{equation} \label{eq:poissonNtrails}
P(N|N_{\max}) = 1 - \left(1 - P(N|\lambda(\lstar_{N}))\right)^{N_{\max}}.
\end{equation}

\section{Example}
\label{sec:examples}

We have applied these techniques to a mock search for \acp{GW} from binary
neutron stars in four days of S5 LIGO data that has been recolored to match the
Advanced LIGO design spectrum~\cite{advLIGO}\footnote{Specifically the
zero-detuned, high-power noise curve was used}.  This provides a potentially
realistic data set that contains glitches from the original LIGO instruments.
A population of neutron star binaries was added at a rate of 4 / Mpc$^3$ / Myr,
(see \citep{rates2010} for the expected rates.)  We self-blinded the signal
parameters with a random number generator. 

Our analysis targeted compact binary systems with component masses between 1.2
and 2 $M_{\odot}$.  We used 3.5 post-Newtonian order stationary phase
approximation templates to cover the parameter space with a 97\% minimal
match~\cite{Owen:1995tm} by neglecting the effects of spin in the waveform
models~\cite{BIOPS}.  This required $\sim$15,000 templates.  We started the
matched filter integrals at 15~Hz and extended the integral to the innermost
stable circular orbit frequency.  The analysis gathered the data, whitened it,
filtered it, identified events in the single detectors, found coincidences and
ranked the events by their joint likelihoods.  The filtering algorithm is
described in~\cite{Cannon2011Early}. 

The previous section described our method for estimating the significance of
events but did not describe many details of how the calculation is done in
practice.  We will point out a few of those details now.

The numerator of \eqref{eq:rank} is evaluated by assuming the signals follow
their expected distribution in Gaussian noise.  We note that this is a
reasonable assumption because detections are likely to come from periods of
relatively stationary and Gaussian data.  Note that the expectation for
$\snr{}$ can be obtained by assuming that sources are distributed uniformly in
space. The expectation for the $\chisq{}$ of a signal can be found
in~\cite{Allen:2004}.
 
The denominator of \eqref{eq:rank} is found by explicitly histogramming the
single detector events that are not found in coincidence.  By excluding
coincident events we lower the chance that a gravitational wave will bias the
noise distribution of the likelihoods.  In general the histogramming will
suffer from finite statistics and ``edge" effects.  We generate the histograms
at a finer resolution than required to track the likelihood and then apply a
Gaussian smoothing kernel with a width characteristic of the uncertainty in
$\snr{}$.

We are unable to collect enough statistics to fully resolve the tail of the
background $\snr{}$ distribution.  Thus, we add a prior distribution into the
background statistics that models the $\snr{}$ falloff as expected from a 2
degree of freedom matched filter in Gaussian noise, i.e.  $p(\rho|n) \propto
\exp{[-\rho^2 / 2]}$.   This helps ensure that the likelihood contours increase
as a function of $\snr{}$ at large $\snr{}$.  At some point the probability of
getting a given value of $\snr{}$, $\chisq{}$ becomes smaller than double
precision float epsilon.  We extend the background distribution above a given
value of $\snr{}$ with a polynomial in $\snr{}$ that falls off faster than the
signal distribution (which is $\propto \rho^{-4}$) but is shallow enough to
prevent numerical problems.  In both cases the point of the prior is not to
influence the ranking of typical events but rather to make the calculations
more numerically well-behaved.  The prior is added so that the total
probability amounts only to a single event in each detector.  Thus the
background (as billions of events are collected) quickly overwhelms the prior
except for at the edges where there is no data.  The point where the
calculation is no longer based on having at least 1 actual event in background
is important since it will effectively mark the limiting \ac{FAP}.  More
discussion of that point follows.

\begin{figure}
\subfigure[]{\includegraphics{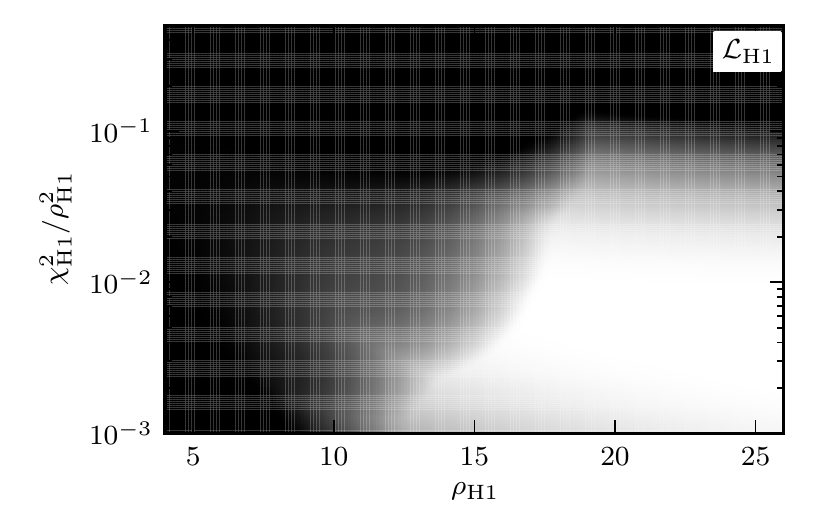}\label{subfig:H1likely}}
\subfigure[]{\includegraphics{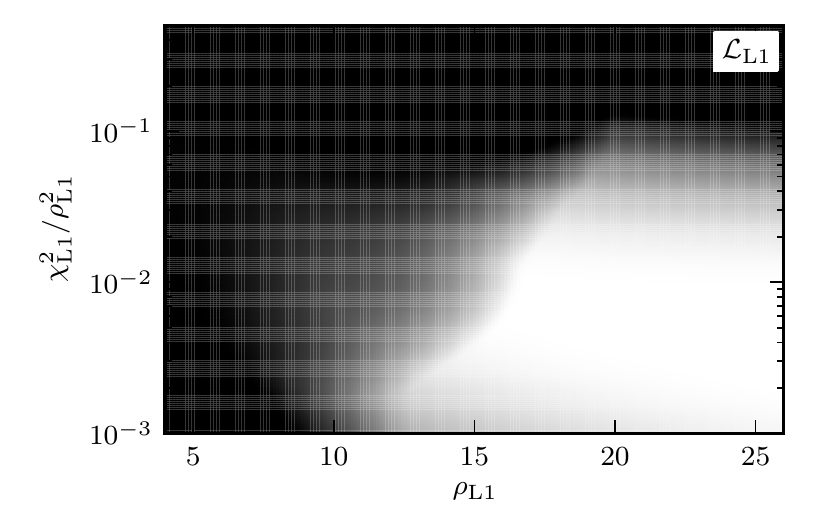}\label{subfig:L1likely}}
\subfigure[]{\includegraphics{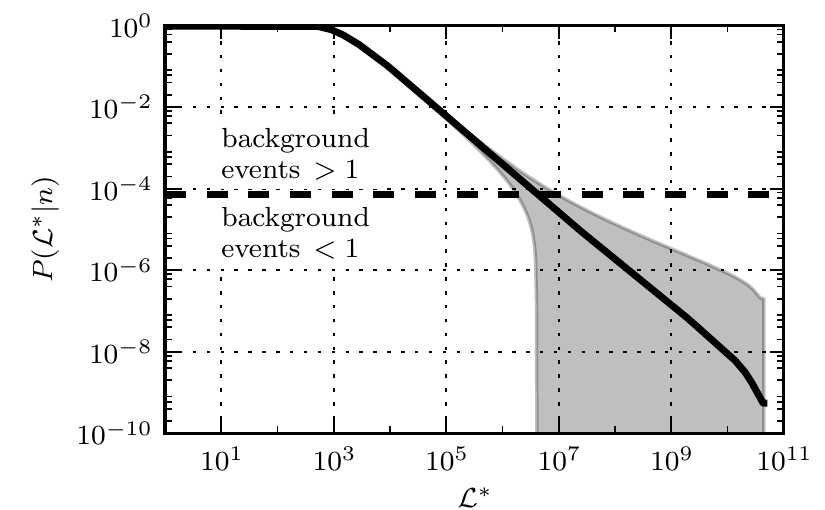}\label{subfig:H1L1likely}}
\caption[]{Figure \ref{subfig:H1likely} and \ref{subfig:L1likely} show the
likelihoods $\likely_{\mathrm{H1}}$,  $\likely_{\mathrm{L1}}$ as a function of
\snr{} and \chisq{} for H1 and L1 respectively for templates with masses
consistent with neutron star binaries (1.2--2 $M_{\odot}$.)
$\likely_{\mathrm{H1}}$, $\likely_{\mathrm{L1}}$ appear as the right-hand-side
of \eqref{eq:LiIndep}.  Lighter colors refer to higher likelihood values.
Figure \ref{subfig:H1L1likely} shows the probability of having obtained a given
value of likelihood $\lstar{}$ or greater from noise as defined in
\eqref{eq:PlstarM} after \trials{} trials (where \trials{} is the number of
independent coincidences formed. In this example $M=6\times 10^{4}$.)  }
\label{fig:likely} \end{figure}
In Fig.~\ref{fig:likely} we show some of the intermediate data used in
estimating the significance of events in our example.  Namely, we show the
individual likelihood contours for $\snr{}$ and $\chisq{}$ described in
\eqref{eq:rank} in the H1 and L1 instruments for signals with a chirp mass
consistent with a neutron star binary ($1.2\, M_{\odot}$) in
Subfig.~\ref{subfig:H1likely} and \ref{subfig:L1likely} respectively.  The
probability of getting an event with a likelihood greater than $\lstar{}$ after
\trials{} trials for the H1 and L1 instruments \eqref{eq:poissonNtrails} is
shown in Fig.~\ref{subfig:H1L1likely}.  Our ability to measure $P(\lstar{}|n)$
is limited by the number of events that we collect in our background estimate.
The shaded region shows the $\sqrt{N}$ error region found by assuming Poisson
errors on the number of events that went into computing a given point on the
curve.  We have indicated the \ac{FAP} at which there ceases to be more than 1
event collected in the background by a dashed line.  The dashed line shows the
$P(\lstar{}|n)$ has background events to $\mathcal{P} := 7\times 10^{-5}$ which
is nearly the \ac{FAP} required for a $4\sigma$ detection.  Below the dashed
line the \ac{FAP} estimate is dominated by the Gaussian smoothing kernel
applied to the planes in Figs.  \ref{subfig:H1likely} and
\ref{subfig:L1likely}.  We believe that it is reasonable to trust the \ac{FAP}
estimate beyond the single background event limit but note that $5\sigma$ level
confidence can still be reached without extrapolation with tighter coincidence
criteria.  Tighter coincidence criteria would reduce the trials factor and
permit higher significances to be estimated.  The best way to do this is to
demand that three or more detectors see an event.  In our example a third
detector would lower the trials factor by $\sim 100$, which would shift the
limiting FAP, $\mathcal{P}$ to $\sim 7 \times 10^{-7}$.  It is worth mentioning
that the background events and number of independent trials are accumulated at
the same rate.  Thus one cannot decrease the limiting \ac{FAP} by collecting
more data.

\begin{figure}
\subfigure[]{\includegraphics{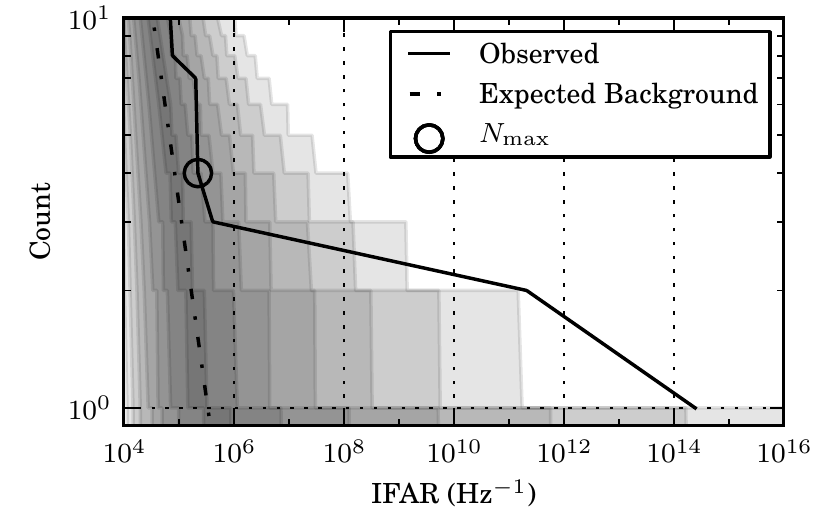}\label{subfig:ifar}}
\subfigure[]{\includegraphics{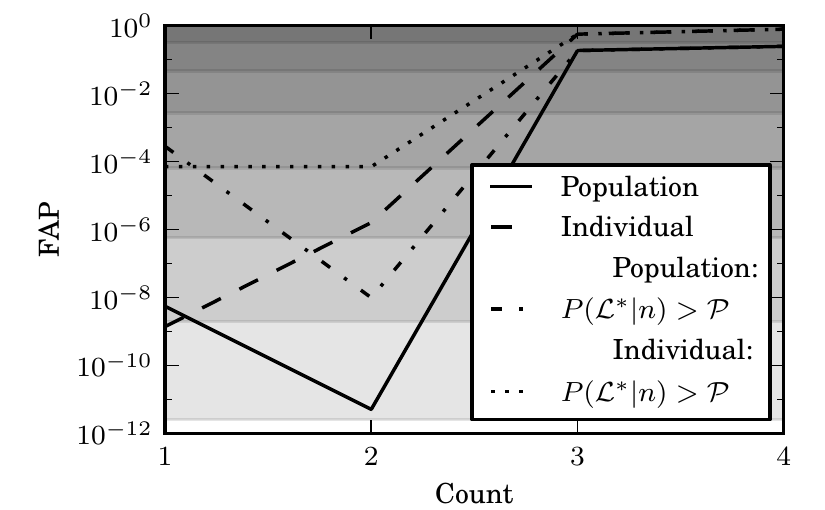}\label{subfig:fap}}
\caption[]{Fig.~\ref{subfig:ifar} is a standard IFAR plot where the shaded
regions correspond to the ``$1\sigma$" through ``$7\sigma$"  regions computed
using the survival function and point percent function associated with the
Poisson distribution. This is used to determine where to stop the accumulation
of events for the population statement.  Fig.~\ref{subfig:fap} shows the
\ac{FAP} associated with each of the individual events in the population we are
considering as well as the \ac{FAP} of obtaining the running $N$ loudest events
without restricting the \ac{FAP} to be greater than $\mathcal{P} = 7\times
10^{-5}$.  Also shown are the same traces obtained after restricting the
\acp{FAP} to be greater than $\mathcal{P}$.}
\label{fig:awesome} \end{figure}
After assigning the \ac{FAP} to events we also assign a \ac{FAR} according to
\eqref{eq:poissonrate}.  This allows us to produce the standard IFAR plot
commonly produced in recent searches for compact binaries~\cite{CBC:S51yr,
CBC:S6} without having relied on time shifting the detector events to estimate
the background.  This is shown in Fig.~\ref{subfig:ifar}.  

The IFARs of the most significant events that came out of this search in
Fig.~\ref{subfig:ifar} can be identified as the long tail in the observed
events distribution. The top event has a significance greater than
$5\sigma$, the level necessary for claiming the detection of \acp{GW}. The
second loudest event has a significance greater than $4\sigma$.  Both events
surpass the single background event limit $\mathcal{P}$. If restricted to this
limit then both events are nearly $4\sigma$.

Applying the population procedure we have put forth in Sec.~\ref{sec:pop}, we
produced a more significant statement about the presence of \acp{GW} beyond
that of the loudest event.  This effect is mostly attributed to the similar
significance of the top two events.  This could happen in a real analysis in
two ways 1) Nature could just provide such a set of events as in this example
2) both events exhaust our ability to measure significance and we must place an
upper bound on the \ac{FAP}.  The latter case, although somewhat artificial,
could still play an important role in analysis, especially if one is unable to
confidently declare a single $5\sigma$ event but finds two or more events with
3 or $4\sigma$.  With our example analysis the combination of the two loudest
events was a $5\sigma$ excursion even after restricting the \ac{FAP} of both
events to be $\mathcal{P}$.  After examining the signal population we found
that both candidates were separately associated with signal injections.

\section{Conclusion}

We have provided a method for estimating the significance of \acp{GW} from
compact binary coalescence using measurements of single instrument populations
of \snr{} and \chisq{} as a function of the template waveform intrinsic
parameters.  We demonstrated our method with mock Advanced LIGO data derived
from initial LIGO data including a realistic population of compact binary
merger signals and glitches.  We found that between our two loudest events we
were able to establish detection  at greater than $5\sigma$ confidence.  Both
of the loudest two events exhausted the $\mathcal{P}$ ($\sim4\sigma$)
background estimate, but the extrapolated \ac{FAP} of the loudest event
exceeded $5\sigma$ on its own. Both of the loudest events were associated with
the blind signal population introduced into the data and the remaining events
were consistent with the expectation from background.

\acknowledgments

Research at Perimeter Institute is supported through Industry Canada and by the
Province of Ontario through the Ministry of Research \& Innovation. KC was
supported by the National Science and Engineering Research Council, Canada.  DK
was supported from the Max Planck Gesellschaft.  This work has LIGO document
number LIGO-P1200031.  The authors wish to thank Rahul Biswas, Patrick Brady,
Gabriela Gonz\'alez Jordi Burguet-Castell, Sarah Caudill and Ruslan Vaulin for
many fruitful discussions.  The authors also thank John Whelan for a careful
review of this work.  We thank the UW-Milwaukee Center for Gravitation and
Cosmology and the LIGO Laboratory for use of its computing facility to make
this work possible through NSF grants PHY-0923409 and PHY-0600953.  We thank
the LIGO Scientific Collaboration and the Virgo Scientific Collaboration for
providing us with the data to test the methods described in this work.

\appendix
\section{Numerical Considerations}

\subsection{Equation \eqref{eq:PlstarM}}

As the duration of the experiment increases, the numerical evaluation of
\eqref{eq:PlstarM} using fixed-precision floating point numbers becomes
challenging.  In this limit, the per-trial false-alarm probability of
interesting events is very small and the number of trials is very large.
Using double-precision floating-point numbers, when the number of trials
gets larger than about \(10^{10}\), \ac{FAP}s of \(10^{-6}\) and 0 become
indistinguishable, and as the number of coincidences that are recorded
increases further ``4\(\sigma\)'' and ``5\(\sigma\)'' events cannot be
differentiated --- it is no longer possible to make detection claims.  The
following procedure can be used to evaluate \eqref{eq:PlstarM} for all
\(P(\lstar|n)\) and \(\trials\).  If \(\trials P(\lstar|n) < 1\) the Taylor
expansion of \eqref{eq:PlstarM} about \(P(\lstar|n) = 0\) converges
quickly.
\begin{multline}
\label{eq:PlstarM:expansion1}
1 - (1 - P)^{\trials}
   = \trials P - (\trials^2 - \trials) \frac{P^2}{2} + \\(\trials^3 - 3
   \trials^2 + 2 \trials) \frac{P^3}{6} - \\(\trials^4 - 6 \trials^3 + 11
   \trials^2 - 6 \trials) \frac{P^4}{24} + \ldots \\
   = \sum_{i = 0}^{\infty} -1^{i} \frac{P^{(i + 1)}}{(i + 1)!} \left[
   (\trials - 0) (\trials - 1) \cdots (\trials - i) \right].
\end{multline}
The last form yields a recursion relation allowing subsequent terms in the
series to be computed without explicit evaluation of the numerator and
denominator separately (which, otherwise, would quickly overflow):  if the
\((i-1)\mathrm{th}\) term is \(X\), the \(i\mathrm{th}\) term in the series
is \(X \frac{i - \trials}{i + 1} P\).

If \(\trials P(\lstar|n) \geq 1\) the Taylor series still converges
(infact, as long as the number of trials \(\trials\) is an integer the
series is exact in a finite number of terms) but the series is numerically
unstable:  the terms alternate sign and one must rely on careful
cancellation of large numbers to obtain an accurate result.  In this regime
the expression's value is close to 1, so \((1 - P)^{\trials}\) is small.  If
\(P\) is small, we can write
\begin{subequations}
\label{eq:PlstarM:expansion2}
\begin{equation}
1 - (1 - P)^{\trials}
   = 1 - \ee^{\trials \ln(1 - P)}
\end{equation}
and then the Taylor expansion of \(M \ln(1 - P)\) about \(P = 0\) converges
quickly,
\begin{equation}
\trials \ln(1 - P)
   = -\trials P \left( 1 + \frac{P}{2} + \frac{P^2}{3} + \ldots \right).
\end{equation}
\end{subequations}

Altogether, the algorithm for evaluating \eqref{eq:PlstarM} is:  if
\(\trials P(\lstar|n) < 1\) use \eqref{eq:PlstarM:expansion1} computed via
the recursion relation;  otherwise if \(P(\lstar|n) < 0.125\) use
\eqref{eq:PlstarM:expansion2};  otherwise evaluate \eqref{eq:PlstarM}
directly using normal floating point operations.  The threshold of
\(P(\lstar|n) < 0.125\) for using \eqref{eq:PlstarM:expansion2} is found
empirically, the results are not sensitive to the choice of this number.

\subsection{Equation \eqref{eq:poissonrate}}

The evaluation of \eqref{eq:poissonrate} for events that are interesting as
detection candidates after an experiment is concluded is straight-forward
using double-precision floating-point arithmetic.  In this regime,
\(P(\lstar|n) \sim 10^{-5}\), and there is plenty of numerical dynamic
range available.  However, the practical use of \eqref{eq:poissonrate} is
in its ability to identify ``once a day'' or ``once an hour'' events for
the purpose of providing alerts to the transient astronomy community.
After just one day, 24 ``once an hour'' background events are expected, and
their \ac{FAP} --- the probability of observing at least one such event
from a Poisson process you expect to have produced 24 --- is
0.9999999999622486.  After 37 events are expected, double-precision numbers
can no longer be used to differentiate those events' \ac{FAP}s from 1;  that
is, \eqref{eq:poissonrate} can only assign reliable false-alarm rates to
the 30 or so most significant background events in any experiment.

This problem is addressed by not computing the expected number of events,
\(\lambda(\lstar)\), from the false-alarm probability, \(P(\lstar)\), as
shown in \eqref{eq:poissonrate}, but by first going back and rewriting
\eqref{eq:cumdist} and \eqref{eq:PlstarM} as
\begin{equation}
1 - P(\lstar|n_1,\ldots,n_{\trials{}})
   = \left( \int_{0}^{\lstar} P(\likely|n) \diff \likely \right)^{\trials},
\end{equation}
from which we can rewrite \eqref{eq:poissonrate} as
\begin{equation}
\lambda(\lstar)
   = -\trials \ln \int_{0}^{\lstar} P(\likely|n) \diff \likely.
\end{equation}
This form of the expression presents no challenges to its evaluation using
double-precision floating point arithmetic.

\bibliography{references}
\end{document}